%% file: 2dxy.tex
\documentclass[12pt]{iopart}

\usepackage{cite}
\usepackage{graphics,epsfig}
\usepackage{graphpap}
\usepackage{subfigure}
\usepackage{lscape}
\usepackage{rotating}

\newcommand{\vect}[2][]{\ensuremath{\mathbf{#2}_{#1}}}

\newcommand{\ave}[2][]{\ensuremath{\left \langle #2^{#1} \right \rangle}}

\newcommand{\vol}[2][]{\ensuremath{\,\mathrm{d}^{#1}{#2}}}

\newcommand{\pdiffb}[2][]{\ensuremath{\frac{\partial^{#1}}{\partial{#2}^{#1}}}}

\newcommand{\infint}{\ensuremath{\int_{-\infty}^{\infty}}}

\newcommand{\npr}[2]{\ensuremath{\ ^{#1}\mathrm{P}_{#2}\ }}

\newcommand{\sumin}[2][0]{\ensuremath{\sum_{{#2}={#1}}^{\infty}}}

\newcommand{\thetabar}{\ensuremath{\overline{\theta}}}

\newcommand{\sumq}[1][]{\ensuremath{\sum_{\vect{q_{#1}}\neq0}}}

\newcommand{\q}[1][]{\ensuremath{\vect[#1]{q}}}

\newcommand{\vr}[1][]{\ensuremath{\vect[#1]{r}}}

\newcommand{\thr}[1][]{\ensuremath{\theta_{\vect{r}{#1}}}}

\newcommand{\dotty}[3]{\put(#1,#2){\circle*{#3}}}

\begin{document}

\title{Temperature dependent fluctuations in the two-dimensional $XY$
model}

\author{S. T. Banks and S. T. Bramwell}

\address{Department of Chemistry, University College London, 21 Gordon
Street, London WC1H OAJ, United Kingdom.}

\ead{simon.banks@ucl.ac.uk}

\begin{abstract}
We present a detailed investigation of the probability density
function (PDF) of order parameter fluctuations in the finite
two-dimensional $XY$ (2d$XY$) model.  
In the low temperature critical phase of this model, the PDF approaches
a universal non-Gaussian limit distribution in the limit $T \rightarrow 0$. 
Our analysis resolves the question of temperature
dependence of the PDF in this regime, for which conflicting results
have been reported. We show analytically that a weak temperature
dependence results from the  
inclusion of multiple loop graphs in a previously-derived graphical
expansion. This is confirmed by numerical simulations on two controlled
approximations to the 2d$XY$ model: the Harmonic and ``Harmonic XY''
models. The Harmonic model has no Kosterlitz-Thouless-Berezinski\u{i}
(KTB) transition and  
the PDF becomes progressively less skewed with increasing temperature
until it
closely approximates a Gaussian function above $T \approx 4\pi$. Near
to that temperature we find some evidence of a phase transition, although
our observations appear to exclude a thermodynamic singularity.
\end{abstract}

\submitto{\JPA}

\section{Introduction}
It is often assumed that global measures of many body systems will be 
normally distributed as a result of the central limit theorem
(CLT).
This requires that the system be separable
into micro- or mesoscopically independent, individually negligible,
elements~\cite{landau_book}. 
When these criteria are not met, the CLT
is not applicable and there is no reason to expect Gaussian
behaviour. A prime example of this comes at equilibrium critical
points where the infinite correlation length violates the condition of
independence.
A universal, non-Gaussian PDF of
a global quantity is thus considered to be a signature of criticality
and the characterization of the forms of these functions
is one of the central problems in the study of critical
phenomena~\cite{wilson_1974_75}. 

The renormalization group assumes that the PDF of a global measure of
a critical system is scale invariant and may thus be obtained from the
appropriate critical fixed
point~\cite{cassandro_1978_913,garrod_book,cardy_book}. 
The statistics
of critical fluctuations are therefore expected to be the same for all
members of a given universality class, as evidenced by studies of
Ising~\cite{bruce_1981_3667,binder_inbook} and
Potts~\cite{botet_2000_1825} models. 
Conversely there is no reason to
assume that systems belonging to different universality classes will
possess the same critical PDFs. 
However in recent years there has been
much interest in the fact that experimental and numerical results from
diverse systems,  
that belong to different universality classes, in many cases
approximate the distribution of the scalar magnetization in the
two-dimensional $XY$ (2d$XY$)
model~\cite{bramwell_1998_552,bramwell_2000_3744}.  
This model supports a continuous line of critical points
constituting a low temperature critical `phase' that is separated from the
paramagnetic region by a Kosterlitz-Thouless-Berezinski\u{i}
transition~\cite{kosterlitz_1974_1046}.
It was observed that the same limit distribution describes the magnetization
fluctuations across all the fixed points on the critical
line~\cite{archambault_1997_8363} 
despite the fact that they belong to different universality classes,
as can be seen from the temperature dependence of the critical
exponents~\cite{archambault_1997_8363,bramwell_2001_041106}. 

The practical and theoretical systems that show PDFs of a functional
form similar to that of the $XY$ model include power fluctuations in
steady state 
turbulence~\cite{bramwell_1998_552,holdsworth_2002_643,bramwell_2000_3744,portelli_2003_104501,peyrard_2004_265,peyrard_2002_834,noullez_2002_231}
and in liquid crystals undergoing electroconvective
flow~\cite{tothkatona_2004_016302,tothkatona_2003_264501,goldburg_2001_245502},
variations in river heights~\cite{bramwell_2002_310}, resistance
fluctuations near electrical
breakdown~\cite{pennetta_2004_s164,pennetta_2004_380},
numerous self-organised critical
systems~\cite{bramwell_2000_3744,sinharay_2001_186,dahlstedt_2001_11193,chapman_2002_409},
dynamical fluctuations in glassy 
systems~\cite{chamon_2004_10120}, and models of equilibrium critical
behaviour~\cite{clusel_2004_046112,zheng_2003_026114}.  
Furthermore, the distribution is related to the Fisher-Tippett-Gumbel
distribution of  
extremal statistics \cite{bramwell_2000_3744} 
which has recently been shown to 
describe the width fluctuations of periodic, Gaussian, $1/f$ noise
\cite{antal_2001_240601}.  
It has been suggested that the apparent universality of the distribution
may be explained by the hypothesis that the low temperature or 
spin wave
limit of the 2d$XY$ model captures the essence of critical behaviour
for a number of universality classes, with any differences that may
occur being hidden by experimental error~\cite{bramwell_1998_552}.
More generally, the low temperature limit of the 2d$XY$ model maps on to the
two-dimensional Edwards-Wilkinson model, a prototypical model of
interface growth. Thus,  
these results also represent an important application of 
interface models to the study of fluctuations in   
complex systems~\cite{racz_1994_3530}.

A team of authors that included one of us analytically derived 
an integral expression for the 
2d$XY$ magnetization PDF in the spin wave regime (sometimes called the
BHP function after reference~\cite{bramwell_1998_552})  
and indeed found it to be temperature-independent in the large-$N$ limit
~\cite{bramwell_2001_041106} . 
This derivation, which applies to the harmonic spin wave approximation
to the $XY$ model in the absence of vortices,  
provides a strong basis for
discussions of the apparently universal form of the PDF. In contrast, 
Palma {\it et al.}  \cite{palma_2002_026108} questioned the result of
temperature independence. In their analysis of Monte Carlo simulations of
the full 2d$XY$ model they found evidence of a weak temperature
dependence of the PDF throughout the low temperature phase.   
However, the full 2d$XY$ model contains both vortex and
anharmonic corrections to the harmonic spin wave model; although
these are considered irrelevant perturbations, that is not to say that
they do not influence the parameters of the PDF in finite
systems. 
It is hard, therefore, to conclude with certainty from numerical
studies of the 2d$XY$ model, that the  
spin wave contribution itself is temperature-dependent. 
The aim of the present work was to establish an analytical basis for
any possible temperature-dependence in the spin wave regime, and to
test it numerically in a controlled manner by means of a series of
approximations to the full 2d$XY$ Hamiltonian.  

The essential properties of the models investigated are summarised
in the following Section (Section 2) . 
Then in  Section 3  we present a further analysis  
of the graphical expansion developed in Ref.~\cite{bramwell_2001_041106}.
Our numerical results are presented in Section 4. Firstly we 
simulate the PDF of the Harmonic model, which lacks both metastable
vortex configurations and anharmonicities, so is directly comparable
with the analytical results. We then simulate the Harmonic XY or HXY
Model, in which vortices are re-introduced into the problem.  
The effect of different definitions of the order parameter is also
investigated. Conclusions are drawn in Section 5.  

\section{The 2d$XY$ Model}
The extended critical region of the 2d$XY$ model obviates the need for
precision control 
of external constraints that one associates with locating an isolated
critical point~\cite{binder_1981_119,binder_inbook,botet_2000_1825}. 
Furthermore the low temperature physics of the model is described
precisely by a harmonic
Hamiltonian~\cite{berezinskii_1971_493,villain_1975_581,archambault_1997_8363}
with the result that many properties are exactly calculable, without
the need for renormalization or the scaling hypothesis.
In the thermodynamic limit the magnetization, $m$, is zero 
for all finite temperatures as a result of the destruction of long
range order by low energy spin waves -- a manifestation of the
Mermin-Wagner theorem~\cite{mermin_1966_1133}. However this limit is
approached sufficiently slowly for $m$ to remain physically relevant
and experimentally
observable~\cite{bramwell_1993_L53,bramwell_1994_8811}. 

The 2d$XY$ model consists of planar spins on a two-dimensional square
lattice with nearest neighbour interactions defined by the Hamiltonian
\begin{equation}
H = -J\sum_{\ave{\vr,\vr[']}}\cos(\thr-\thr[']).\label{eq:cos_ham}
\end{equation}
Here $J$ is the (ferromagnetic) exchange interaction, $\thr$ is the
angle (relative to some arbitrary but
fixed axis) made by the spin located at $\vr$ and the sum runs over
all nearest neighbour pairs of 
spins. We always deal with a periodic square lattice with sides $L$,
such that $N=L^2$. 

Renormalization below the Kosterlitz-Thouless-Berezinski\u{i} transition temperature 
removes vortices from the
problem~\cite{cardy_book,jose_1977_1217} and the model is
described by the Harmonic model
Hamiltonian~\cite{berezinskii_1971_493} 
\begin{equation}
H = J\left[1-\frac{1}{2}\sum_{\ave{\vr,\vr[']}}(\thr-\thr['])^2\right].
\label{eq:harmonic}
\end{equation}
This presents a difficulty with regard to the periodicity of the spin
variables implicit in (\ref{eq:cos_ham}). The average product of two
spin variables defines a Green's function analogous to a propagator
of a Euclidean free field. This should be able to take the full range
of values between $\pm\infty$ and so the spins in this model are
necessarily non-periodic~\cite{berezinskii_1971_493}. 

Intermediate between the Harmonic and full 2d$XY$
models is the H$XY$ model~\cite{archambault_1997_8363}, defined by 
\begin{equation}
H = \frac{J}{2}\sum_{\ave{\vr,\vr[']}}(\thr-\thr[']-2n\pi)^2.
\end{equation}
This is numerically simpler than the Villain
model~\cite{villain_1975_581} as the parameter $n$ is restricted to
the values 
$0,\pm1$ to ensure that the contents of the brackets remain bounded by
$\pm\pi$. As in the full 2d$XY$ model, the spins are defined modulo $2\pi$.

The quantity we study is the
instantaneous scalar magnetization, related to the vector order
parameter by $m=|\vect{m}|$. One definition makes use of the
vector spins,
\begin{equation}
m =
\frac{1}{N}\sqrt{\left(\sum_{\vr}\vect[\vr]{s}\right)^2} .
\label{eq:op_vect}
\end{equation}
We refer to this as the `full' order parameter. A more analytically
useful expression defines the magnetization in terms of the
spin variables,
\begin{equation}
m =
\frac{1}{N}\sum_{\vr}\cos\psi_{\vr}\label{eq:op_cos} \hspace{1cm}
\mathrm{with} \hspace{1cm} \psi_{\vr}=\thr-\thetabar.
\end{equation}
where \thetabar\ is the instantaneous magnetization direction. For
unconfined spins this is the average, $\ave{\thr}$, for a given
configuration. For periodic spins it is convenient to
define~\cite{archambault_1997_8363} 
\begin{equation}
\thetabar = \frac{\sum_{\vr}\sin{\thr}}{\sum_{\vr}\cos{\thr}},
\end{equation}
which is the same as the mean for large $N$. Defining $\psi$ enables
one to work in a reference frame in which the Goldstone mode has been
removed. 

At low temperatures angular differences between spins are likely to be
small and it may be possible to truncate the cosine in
(\ref{eq:op_cos}). This defines the so-called `linearized' order
parameter,
\begin{equation}
m = 1 - \frac{1}{2N}\sum_{\vr}\psi_{\vr}^2.\label{eq:op_lin}
\end{equation}

\section{Evaluation of $P(m)$}
We are interested in the critical order parameter fluctuations of finite
systems which we discuss in terms of probability density functions (PDFs),
$P(m)$, calculated in the thermodynamic limit. Thus the order
parameter must be correctly normalized to avoid the width of the PDF
becoming either zero or infinite as $N\rightarrow\infty$. This has
been extensively 
discussed elsewhere~\cite{bramwell_1998_552,bramwell_2001_041106} and
requires that $m$ be scaled by the standard deviation. We define
\begin{equation}
\Pi(z) = \sigma P(m) \hspace{2cm} z=\frac{m-\ave{m}}{\sigma}.
\end{equation}

\subsection{Evaluation of the Moments $\ave[p]{m}$}
A distribution function may be defined in terms of its moments
as~\cite{kendall_book} 
\begin{equation}
P(m) = \infint\frac{\vol{x}}{2\pi}\rme^{\rmi mx}
\sumin{p}\frac{\left(-\rmi x\right)^p}{p!}\ave[p]{m}.\label{eq:Pm}
\end{equation}
Following from the definition of the order parameter in equation
(\ref{eq:op_cos}),
\begin{equation}
\ave[p]{m}  = 
\frac{1}{(2N)^p}\mathrm{Tr}\ave{\exp\left(\sum_{a=1}^p\rmi\sigma_a\psi_{\vr[a]}\right)\!}.
\label{eq:mp1}
\end{equation}
where $\rmi$ is the imaginary unit and the trace is over all $\vr[i]$
from 1 to $N$ and all $\sigma_i=\pm1$.
Using Gaussian integration ($\psi_{\vr}$ is a Gaussian
variable) we can derive~\cite{archambault_1998_7234}
\begin{equation}
\ave[p]{m}  =  \left(\frac{\ave{m}}{2N}\right)^p
\sum_{k=0}^\infty\left\{\frac{1}{k!}\left(-\frac{\tau}{2}\right)^k
\Tr\left[\left(\sum_{a\neq 
b}^p\sigma_aG_{ab}\sigma_b\right)^k\right]\right\}\label{eq:mp}
\end{equation}
In this expression $\tau=T/J$, $G_{ab}$ is the Green's function propagator
function \mbox{$\tau G(\vr[a]-\vr[b])=\ave{\psi_{\vr[a]}\psi_{\vr[b]}}$},
$(a,b)$ represents all possible pairs of $a$ and $b$ and the
shorthand $\sum_{a\neq b}=\sum_{a}\sum_{b\neq a}$ has been introduced. 
The upper limits are included explicitly as a reminder that
the sums run over $p$ spins, not the entire lattice.

\subsection{A Graphical Representation of the Moments}
Equation (\ref{eq:mp}) is exact. To proceed further one must develop
a means 
of expressing the sums so that they may either be evaluated precisely,
or approximated in a controlled manner. We
follow~\cite{bramwell_2001_041106} and consider a graphical
representation of the moments which allows us to perform well defined
partial summations.
It is possible to interpret equation (\ref{eq:mp}) diagrammatically
by letting 
$G_{ab}$ represent a line joining two distinct points $a$ and $b$ on
a sublattice of $p$ points chosen from the original lattice. 
The $k^{\mathrm{th}}$ term in the argument of the trace is the
sum of contributions from all possible graphs with $k$ lines on a
sublattice of $p$ 
points chosen from a parent lattice of size $N$.

\begin{figure}
\begin{center}
\begin{minipage}[t]{0.4\linewidth}
\centering
\subfigure[]
{\input{figs/graphs_a}
\label{fig:graphs_a}}
\end{minipage}
\begin{minipage}[t]{0.4\linewidth}
\centering
\subfigure[]
{\input{figs/graphs_b}
\label{fig:graphs_b}}
\end{minipage}
\begin{minipage}[t]{0.4\linewidth}
\centering
\subfigure[]
{\input{figs/graphs_c}
\label{fig:graphs_c}}
\end{minipage}
\begin{minipage}[t]{0.4\linewidth}
\centering
\subfigure[]
{\input{figs/mlgs}
\label{fig:graphs_d}}
\end{minipage}
\caption[Graphs in the Expansion of Moments of $m$]{Graphs
in the Expansion of \ave[p]{m}: (a) Examples of 
allowed connected graphs with $p=2,3,4$ and $k=p$. The graphs in (b)
are disallowed as 
they contain either odd vertices or loops involving a single
point. (c) An example of a disconnected graph (left hand side) which 
makes the same contribution as the product of its constituent
connected parts. (d) A multiple loop graph.}
\label{fig:graphs}
\end{center}
\end{figure}
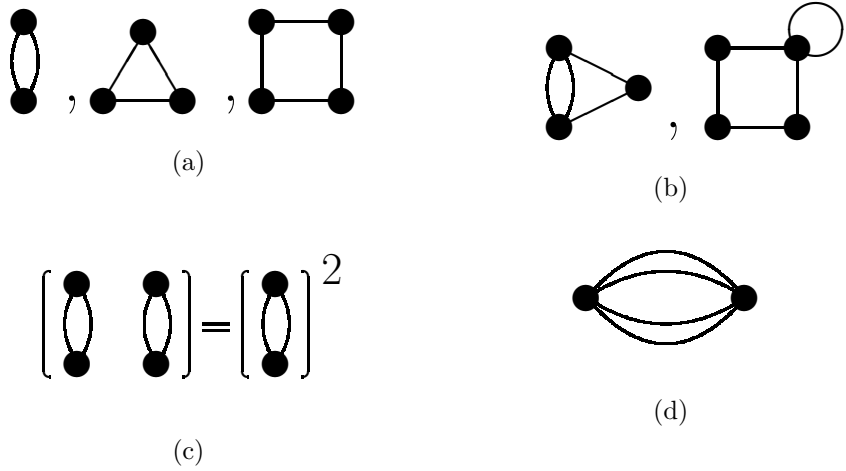

Graphs with lines beginning and ending at the same point are
disallowed (by virtue of the constraint $a\neq b$), and graphs with
odd vertices (points with an odd number of connections) contribute zero
after performing the trace. Contributions from disconnected graphs
(those for which the lines do not form a continuous path, e.g.\
\fref{fig:graphs_c}) are 
equal to the product of their connected constituents.

Relevant graphs belong to one of two categories: `Single Loop Graphs'
(SLGs) contain only points with no 
more than two lines attached; `Multiple Loop Graphs' (MLGs) may have
points with any (even) number of lines attached.  
It has previously been assumed~\cite{bramwell_2001_041106} that MLGs
do not contribute to the moments of $m$ in the thermodynamic
limit. We begin by reviewing the SLG only approach, to show
how this yields a 
temperature independent form of $\Pi(z)$, before returning to consider
the validity of neglecting MLGs.

\subsubsection{The Value of Single Loop Graphs}
Evaluation of (\ref{eq:mp}) requires knowledge of the value of each
type of graph and the number of different ways each graph can be
created. 
For connected SLGs it is always possible to rearrange the order of the
$G_{ab}$ so that they form a simple closed loop. Transforming to
Fourier space using $G(\vr) = (1/N)\sumq\rme^{\rmi\q.\vr}G(\q)$ then gives
an expression for the value of the trace in (\ref{eq:mp}),
\begin{eqnarray}
\fl\Tr\left[\left(\sum_{a\neq 
b}^p\sigma_aG_{ab}\sigma_b\right)^k\right] & = &
C(k)\frac{2^p}{N^k}\sum_{\vr[1]=1}^N\cdots\sum_{\vr[p]=1}^N
\sum_{a=1}^p\sum_{b\neq a}^p\cdots\sumq[1]
\rme^{\rmi\q[1].(\vr[a]-\vr[b])}G(\q[1])\nonumber\\
&  & \times \sumq[2]\rme^{\rmi\q[2].(\vr[b]-\vr[c])}G(\q[2]) 
\cdots \sumq[k]\rme^{\rm i\q[k].(\vr[k]-\vr[1])}G(\q[k])\\
& = & C(k)2^{p}\npr{p}{k}N^{p}g_k.\label{eq:S'}
\end{eqnarray}
The factor $C(k)=2^{k-1}(k-1)!$ accounts for repetition
of topologically identical graphs. The $g_k$ are numerical constants
defined by $g_k=(1/N)^k\sumq(1/\gamma_{\q})^k$ with
$\gamma_{\q}=G(\q)^{-1}=(4-2\cos q_x-2\cos q_y)^{-1}$. Evaluation of these
constants is discussed in appendix B of
reference~\cite{bramwell_2001_041106}\footnote{We note a correction to
equation (B3) in reference~\cite{bramwell_2001_041106}. The correct
expression for $g_k$ is
\begin{displaymath}
g_k =
\frac{1}{4^{k-1}\pi^{2k}}\left\{\sum_{x=1}^{\infty}\sum_{y=1}^{\infty}
\frac{1}{(x^2+y^2)^k}+\sum_{x=1}^{\infty}\frac{1}{x^{2k}}\right\}.
\end{displaymath}
The numerical values that appear elsewhere in that paper are accurate
so the error appears to be typographical.}

Adjusting the factor $C(k)$ to account for the extra combinatorial
considerations imposed by disconnected graphs leads
to~\cite{bramwell_2001_041106}
\begin{equation}
\frac{\ave[p]{m}}{\ave{m}^p} 
 = 
\exp\left(\sum_{k=2}^\infty\frac{g_k}{2k}(-\tau)^k\left.\pdiffb[k]{z}z^p\right|_{z=1}\right).
\label{eq:mp6}
\end{equation} 
The standard deviation is thus $\sigma = \sqrt{{g_2}/{2}}\tau\ave{m}$,
from which it is seen that hyperscaling
is obeyed~\cite{bramwell_2001_041106,bramwell_2000_3744}. Substituting
(\ref{eq:mp6}) into (\ref{eq:Pm}) and making the 
transformation $x\rightarrow x/\sigma$ gives the BHP
distribution~\cite{bramwell_2001_041106,bramwell_2000_3744}, 
\begin{equation}
P(m) = \infint \frac{\vol{x}}{2\pi\sigma}\exp
\left\{\frac{\rmi x(m-\ave{m})}{\sigma} + \sum_{k=2}^\infty\frac{g_k}{2k}
\left(\rmi x\sqrt{\frac{2}{g_2}}\right)^k\right\},\label{eq:bhp}
\end{equation}
which is explicitly independent of both system size and
temperature. \Eref{eq:bhp}, obtained using an SLG only approach, is
the main result of~\cite{bramwell_2001_041106}.

\subsection{The Effect of Introducing MLGs}
\label{sec:mlgs}
The simplest MLG consists of a sublattice with $p$ points, only two of
which are connected by $k>2$ 
lines (e.g.\ Figure~\ref{fig:graphs_d}). Restricting the trace to only
this type of graph and again transforming to reciprocal space gives
\begin{equation}
\Tr\left[\left(\sum_{a\neq 
b}^p\sigma_aG_{ab}\sigma_b\right)^k\right] 
 =  2^{p+k-1}N^p\npr{p}{2}\Theta_k\label{eq:SMLG}
\end{equation}
where
\begin{equation}
\Theta_k = \frac{1}{N^k}\sum_{\q[1]\neq0,\ldots,\q[k]\neq0}G(\q[1])
\ldots G(\q[k])\delta\left(\sum_i \q[i]\right).\label{eq:thetak}
\end{equation}
The symmetry of $G(\q)$ means that $\Theta_2=g_2$ and so, considering
only the additional (MLG) contributions from the highly specific set of
graphs discussed here, the second moment becomes
\begin{equation}
\ave[2]{m} = \ave{m}^2 \left(1+g_2\tau^2+\frac{1}{24}\Theta_k
\tau^4+\ldots+\right).
\end{equation}
Therefore in the low
temperature limit only the single loop graphs are significant as all
MLGs correspond to higher powers of $T$, however, for high temperatures,
contributions from MLGs may not be ruled out.

Substituting $G(\q)=1/\gamma_{\q}$ into (\ref{eq:thetak}), together
with the approximation $\gamma_{\q}\approx \q^2$ (representing the
greater weight of the low frequency modes), yields,
after a little algebra
\begin{eqnarray}
\fl \Theta_k = \frac{1}{(2\pi)^{2k}}
\sum_{\{x_i\}=-\infty}^{\infty}\sum_{\{y_i\}=-\infty}^{\infty}
\frac{1}{(x_1^2+y_1^2)}\frac{1}{(x_2^2+y_2^2)}
\times\ldots\nonumber\\ 
\times\frac{1}{(x_{k-1}^2+y_{k-1}^2)}
\frac{1}{\left((-\sum_{i=1}^{k-1}x_i)^2+(-\sum_{i=1}^{k-1}y_i)^2\right)}.
\end{eqnarray}
The sum over $\{x_i\}$ is used to indicate a multiple sum over all
members of the set $\{x_1,x_2,\ldots,x_k\}$. The terms in this sum are
all positive and therefore cannot cancel each other.
From this we can see that
$\Theta_k$ is explicitly non-zero: thus the contributions from MLGs may
not be neglected in the thermodynamic limit, contrary to
reference~\cite{bramwell_2001_041106}. 

\section{Monte Carlo Simulations}

\subsection{Simulations of the Harmonic Model}
\label{sec:2dXY_sim_harm_model}
The Harmonic model (\ref{eq:harmonic}) describes
the physics of the critical region of the 2d$XY$
model~\cite{kosterlitz_1974_1046,jose_1977_1217}. Thus if $\Pi(z)$ is
independent of $T$~\cite{bramwell_2001_041106} the order parameter
fluctuations in the Harmonic model should be BHP like at all
temperatures. Single spin-flip Metropolis~\cite{metropolis_1953_1087}
Monte Carlo simulations were performed on 
Harmonic systems with $L=10,12,14\ldots32$, over a range of
temperatures from $T/J=0.5$ to 
as high as $T/J=50$. Equilibration was over $10^6$ Monte Carlo steps
per spin (MCS/s) with $10^7$ MCS/s in total.

\begin{figure}
\begin{center}
\begin{minipage}[t]{0.45\linewidth}
\centering
\subfigure[]
{\rotatebox{-90}{\epsfig{figure=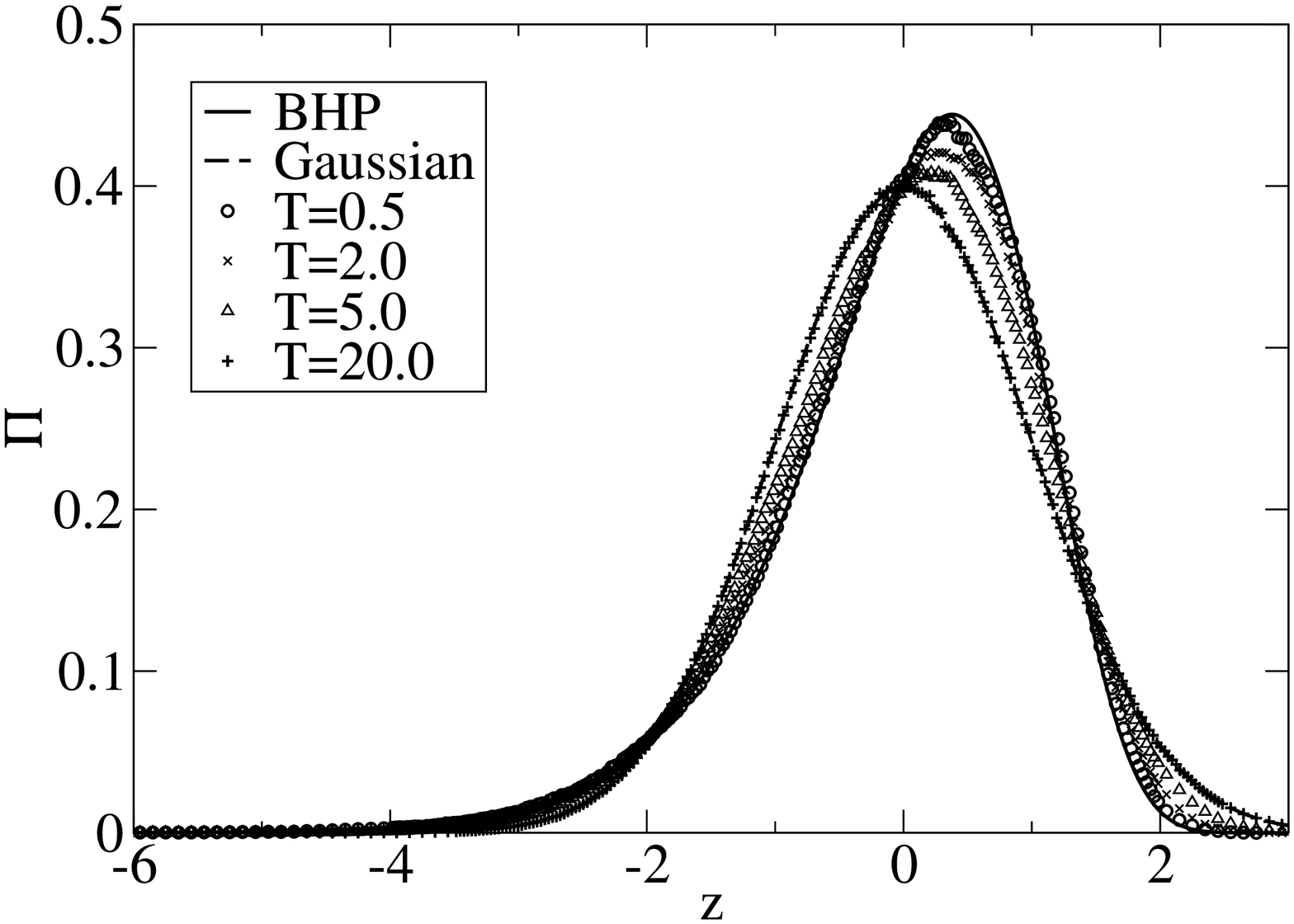,width=0.8\linewidth}}
\label{fig:mc_harmonic}}
\end{minipage}
\begin{minipage}[t]{0.45\linewidth}
\centering
\subfigure[]
{\rotatebox{-90}{\epsfig{figure=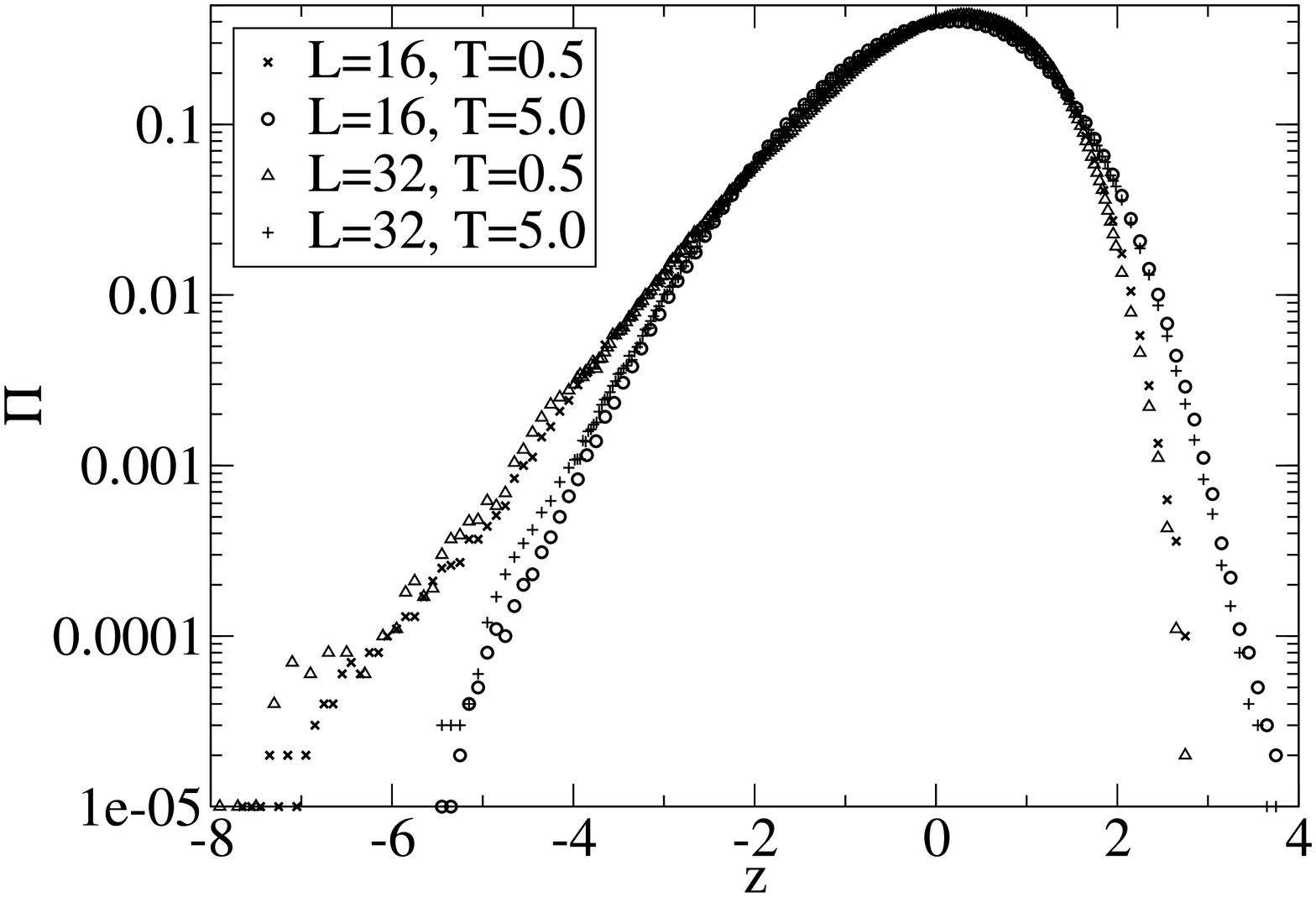,width=0.8\linewidth}}
\label{fig:Nind}}
\end{minipage}
\caption[Monte Carlo Simulations on the Harmonic Model]{Monte Carlo
Simulations on the Harmonic Model. (a) $L=32$: There is a clear
variation with temperature from an approximately BHP form at $T=0.5$
to a perfect Gaussian by the time $T=20.0$. (b) Our simulations showed
effectively no dependence on system size for $N>100$. These results
are for $L=16$ and $L=32$ at $T/J=0.5,5.0$. $\Pi$ and $z$ are as
defined in the text.}
\end{center}
\end{figure}

The results (figure~\ref{fig:mc_harmonic}) show a clear variation of
$\Pi(z)$ with $T$ (for clarity only the data for $L=32$ is shown). At
low  
temperatures the distribution is excellently described by the BHP
function. However as $T$ increases the PDF becomes progressively less
skewed, eventually becoming Gaussian. This is as expected in the light
of the work presented in section~\ref{sec:mlgs}. Higher powers of $T$
enter the moment expansion as a result of the inclusion of MLGs. For
$r>2$, $\sigma^r$ dominates the other moments at high $T$ and the
normalized cumulants tend to zero. We observed no variation in the
form of the PDF as a function of system size at any temperature, as
highlighted in figure~\ref{fig:Nind}. 

\begin{figure}
\begin{center}
\begin{minipage}[t]{0.45\linewidth}
\centering
\subfigure[]
{\rotatebox{-90}{\epsfig{figure=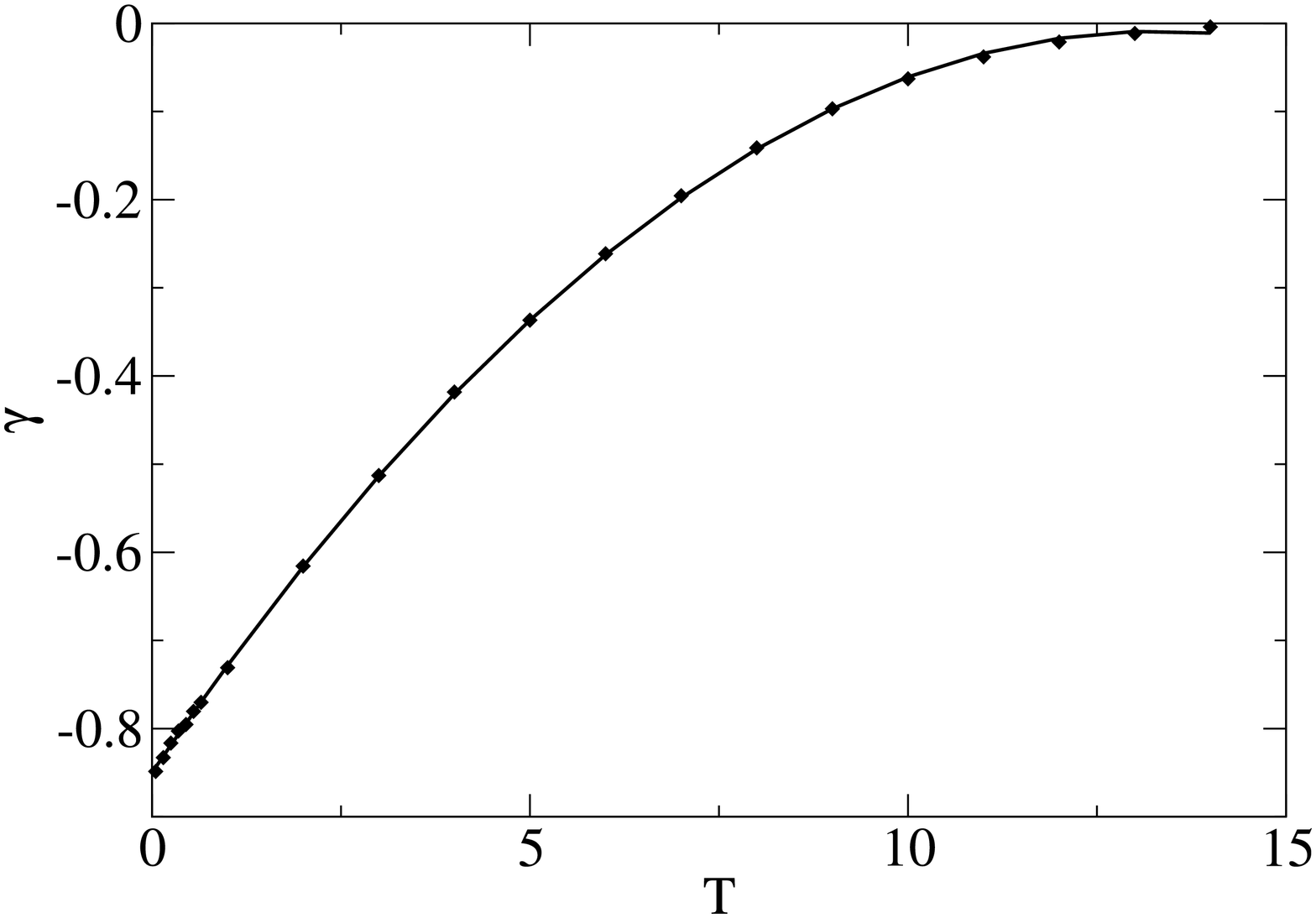,width=0.8\linewidth}}
\label{fig:skew_harmonic}}
\end{minipage}
\begin{minipage}[t]{0.45\linewidth}
\centering
\subfigure[]
{\rotatebox{-90}{\epsfig{figure=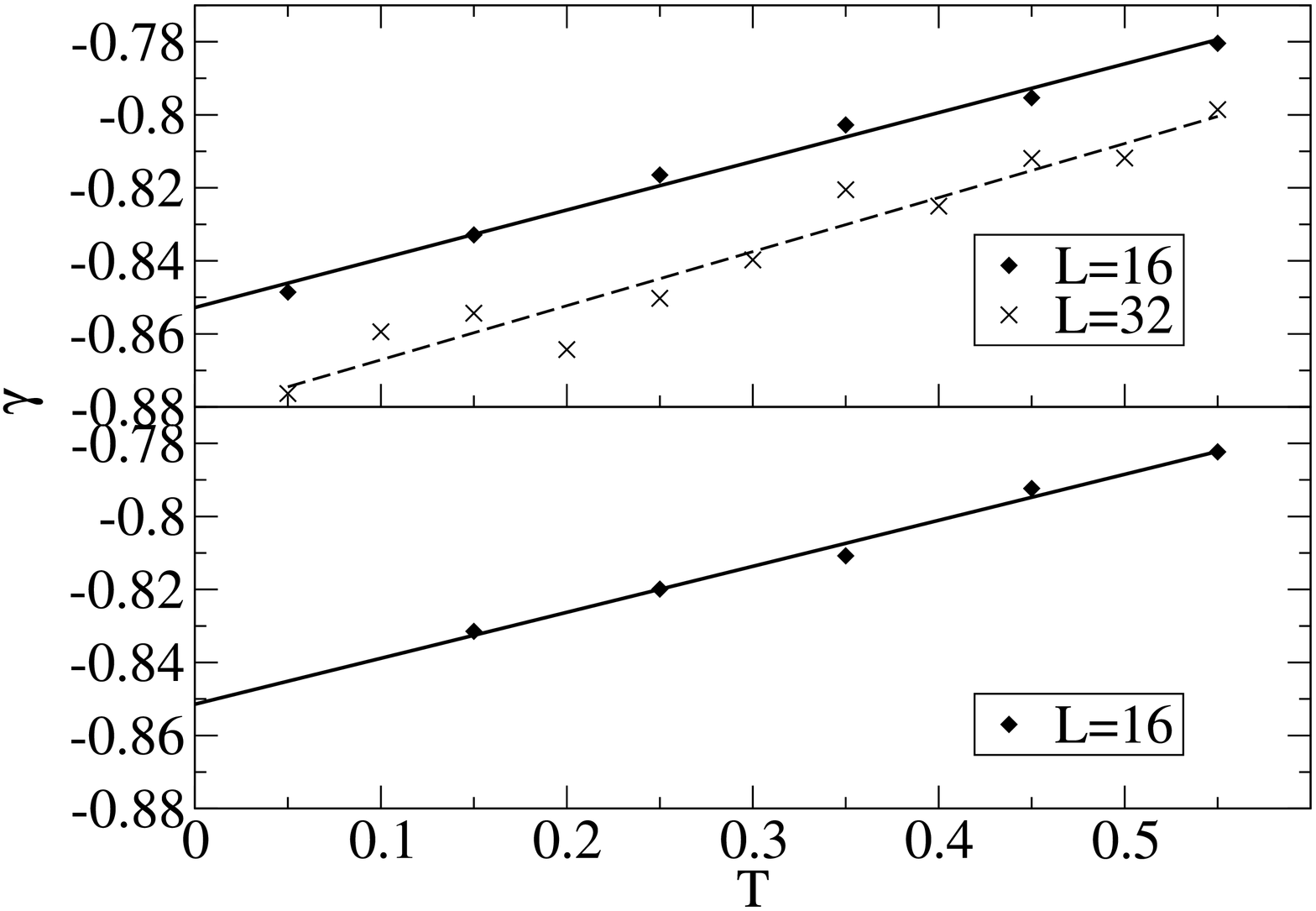,width=0.8\linewidth}}
\label{fig:skew_linear_harmonic_HXY}}
\end{minipage}
\renewcommand{\baselinestretch}{1}
\caption{Temperature dependence of the skewness. (a) The Harmonic
model with $L=16$ over a wide range of temperatures. (b) The low
temperature region for the Harmonic model (upper plot) with $L=16$
(diamonds) and $L=32$ (stars) and the H$XY$ model (lower plot) with
$L=16$, confirming that they obey approximately the same linear  
relationship in the critical region. The solid and dashed lines are:
(a) a least 
squares quadratic fit; (b) linear regression curves; equations in the
text.} 
\end{center}
\end{figure}

The skewness, $\gamma_3$, provides a clear measure of the variation of
the PDF with temperature. We have confirmed that $\gamma_3$ is
independent of $N$ for large systems (though minor corrections to
this universality are observed for small $N$) and a least squares fit
to the numerical data (figure~\ref{fig:skew_harmonic}) reveals that
the steady decrease in $|\gamma_3|$ can be approximated (for $L=16$)
by (taking $J=1$)
\begin{eqnarray}
\gamma_3(T) & \approx & -0.85+0.126T-0.0048T^2\\
& \approx & -0.85 + 0.79\eta-0.19\eta^2,
\end{eqnarray}
up to the point at which the skewness becomes zero at around
$\eta\approx2$ (here $\eta=T/(2\pi)$ is the spin wave exponent). The
zero temperature value of $\gamma_3=-0.85$ is 
slightly higher than the theoretical BHP value of $\gamma_3=-0.89$. We
attribute this to the relatively small system used and show
in figure~\ref{fig:skew_linear_harmonic_HXY} that increasing the size
to $L=32$ gives a limiting value much closer to the theory with
$\gamma_3(T)\approx -0.88+0.15T$. We fully expect the true universality
with respect to $N$ to be evident for larger systems. However, the
skewness is relatively computationally expensive due to the need for
averaging 
and as our results do confirm the evolution of $\gamma_3$ with $T$ we
leave the determination of the the precise form of $\gamma_3(T)$ from
larger systems to another time.

\subsection{Temperature Dependence in the H$XY$ Model}
\label{sec:mc_HXY}
Renormalization of the H$XY$ model below $T_{KT}$ removes all vortices
and recovers the Harmonic model at some, generally higher,
temperature. Using the renormalization 
procedure described by Jos\'e \emph{et al}.~\cite{jose_1977_1217} we
have calculated the mapping of temperature scales between these two
models (table \ref{table:rgdata}) showing that they are coincident
below $T\leq0.9$. 

Simulations of the H$XY$ model in this range show a clear change in
the skewness of the magnetization 
distribution as $T$ is varied. Given that there are effectively no
vortices present in the system, it is concluded that the temperature
dependence has the same origins as in the Harmonic model, namely the
multiple loop graphs in the moment expansion. 
\begin{table}
\centering
\input{figs/data}

\renewcommand{\baselinestretch}{1}
\caption[Equivalent Temperature Scales for the H$XY$ and Harmonic
Models with \mbox{$L=16$}]{Equivalent Temperature Scales for the H$XY$
and Harmonic 
Models with \mbox{$L=16$}: The temperature of the H$XY$ model is given as
$T$. $T_{\mathrm{eff}}$ and $K_{\mathrm{eff}}$ are the effective
temperature and spin wave stiffness respectively -- i.e.\ relating to
the Harmonic model in which all vortices have been renormalized
out. The RG 
expansion parameter $y$ is a measure of the vortex density.}
\label{table:rgdata}
\end{table}
The plot of $\gamma_3(T)$ for the harmonic model
(figure~\ref{fig:skew_harmonic}) shows an essentially
linear dependence at low temperatures. We confirm that a very similar
form is observed for the H$XY$ model
(figure~\ref{fig:skew_linear_harmonic_HXY}), both $L=16$ models adhering to 
\begin{equation}
\gamma_3 \approx 0.82\eta-0.85.
\end{equation} 

\subsection{The Effect of the Order Parameter}
\label{sec:2dxy_op_diffs}
To enable direct comparison with the theoretical predictions all the
simulations discussed above have used the cosine form of the order
parameter, equation (\ref{eq:op_cos}). It is interesting to consider
the effect 
on the Harmonic model
of substituting for this either the full order parameter, or
the much discussed~\cite{bramwell_2001_041106} linearized form
(\ref{eq:op_lin}). 

\subsubsection{The Full Order Parameter}
\begin{figure}
\begin{center}
\rotatebox{-90}{\epsfig{figure=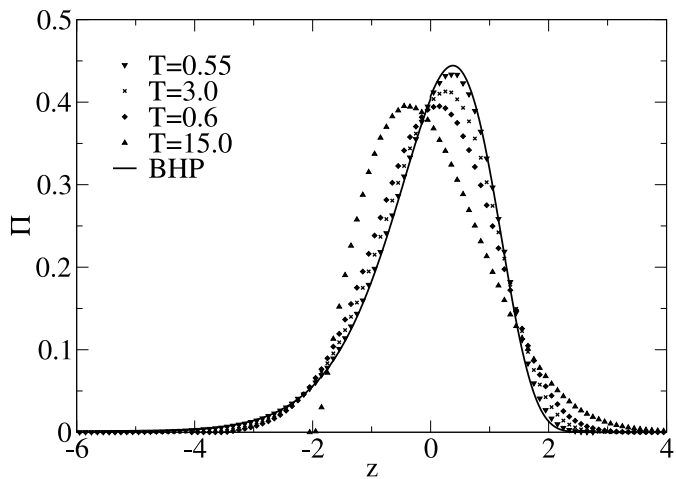,width=0.5\linewidth}}
\renewcommand{\baselinestretch}{1}
\caption[Monte Carlo Results for the Harmonic Model with Full Order
Parameter ($L=16$)]{Monte Carlo Results for the Harmonic Model with
Full Order 
Parameter ($L=16$): Again there is a clear variation of the
distribution as a function of temperature. At low $T$ the curve
closely approximates the 
BHP form, but with the full order parameter it becomes Maxwellian,
rather than Gaussian, at high $T$ (c.f.\ figure~\ref{fig:mc_harmonic}).}
\label{fig:2dXY_harm_vect}
\end{center}
\end{figure}

High temperature ($T>T_{KT}$) simulations of the 2d$XY$ and H$XY$
models with the cosine order parameter lead to Maxwellian rather than
Gaussian magnetization 
distributions. We suggest that this difference from the Harmonic model is
not a consequence of vortices but arises from the constraints placed
upon the spin degrees of freedom. The periodicity of the spins forces
$m$ to be positive and the PDF is a slice through a two-dimensional
Gaussian, appearing as a Maxwell distribution of speeds. A similar
result may be obtained for the harmonic model by using the full order
parameter (\ref{eq:op_vect}). This quantity is necessarily positive
and thus has the same effect as 
constraining the spin variables to within the positive range of the 
cosine function in (\ref{eq:op_cos}). Monte Carlo simulations confirm
this, showing BHP 
fluctuations changing to Maxwellian behaviour as $T$ is increased
(figure~\ref{fig:2dXY_harm_vect}). Despite the Harmonic model/full
order parameter combination giving the same limiting $T$ PDFs as the
2d$XY$ and H$XY$ models, it should be noted that the path of
intermediate distributions is different and only for the Harmonic
model is there a region of Gaussian behaviour.

\subsubsection{The Linearized Order Parameter}
\label{sec:op_quad}
The linearized order parameter (\ref{eq:op_lin}) only describes $m$ at
temperatures low 
enough to exclude large angular differences between neighbouring
spins. Despite this, Bramwell 
\emph{et al}. demonstrated that, for the Harmonic 
model, the distribution of the linearized order parameter is precisely
the BHP function at all temperatures~\cite{bramwell_2001_041106}. We
have simulated this model for a system with $N=1024$ and the results
confirm that the magnetization distribution is consistently of the BHP
form for all our simulations, even up to $T/J=50$.
This result may now be understood as a manifestation of the low
temperature approximation implicit in the derivation of the BHP
distribution (\ref{eq:bhp})~\cite{bramwell_2001_041106}. The neglect
of MLGs leads to a PDF which perfectly describes critical fluctuations
in the 2d$XY$ model, but strictly only in the limit
$T\rightarrow0$. The linearized OP is only a valid description of the
magnetization in the same limit and hence has the same temperature
independent fluctuations as the full OP at $T=0$. We
emphasize that we have not demonstrated an analytical equivalence
between the neglect of MLGs and the neglect of anharmonic terms in the
cosine OP. Despite this it seems likely that this heuristic
justification of the rigorous equivalence between the PDFs obtained in
each case is correct.

\section{Is There A Phase Transition in the Harmonic Model?}
The change from non-Gaussian to Gaussian statistics in our Harmonic
model simulations is reminiscent of a Kosterlitz-Thouless-Berezinski\u{i} transition
from a critical region to a paramagnetic
phase~\cite{kosterlitz_1973_1181}. The temperature at which
Gaussianity is first observed ties in with the expression for the 
susceptibility derived in~\cite{archambault_1997_8363} 
\begin{equation}
\chi =\frac{N\ave{m}^2T}{2a_{2d}J^2} \sim N^{1-T/4\pi J}\label{eq:chi}
\end{equation}
($a_{2d}\approx 258.6$) which has
$\chi$ diverging only for $T/J<4\pi$. There is, however, no 
change in the behaviour of the magnetization as one passes through
this point as one would expect for a true phase change and it must be
remembered that (\ref{eq:chi}) is only
approximate~\cite{archambault_1997_8363}.  

\begin{figure}
\begin{center}
\subfigure[]{
\framebox{\epsfig{figure=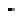,width=0.25\textwidth}}}
\subfigure[]{
\framebox{\epsfig{figure=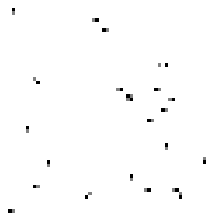,width=0.25\textwidth}}}
\subfigure[]{
\framebox{\epsfig{figure=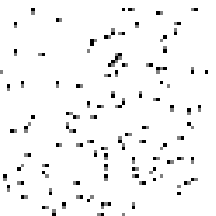,width=0.25\textwidth}}}
\renewcommand{\baselinestretch}{1}
\caption[Vortices in the Harmonic Model]{Vortices in the Harmonic
Model: These figures are snapshots 
of the Harmonic model with $N=3600$ at (a) $T/J=1.5$, (b) $T/J=2.5$,
and (c) $T/J=3.5$. At low temperatures
one sees the surprising emergence of a tightly bound
vortex/anti-vortex pair. The vortex density increases with $T$ and
despite a small degree of separation of some pairs in (b) and (c),
isolated vortices are not observed. The grey squares represent a
vorticity of -1, the black squares a vorticity of +1.} 
\label{fig:2dXY_harmonic_vortices}
\end{center}
\end{figure}

The KT transition is associated with the onset of topological order,
which suggests that, if such a transition were to occur in the
Harmonic model, a suitable topological defect must be found. We
have identified one such defect, and, rather surprisingly, it is a spin
vortex. The definition of a vortex is complicated by the
non-periodicity of the spins. We assign a `winding number' to each
pair of spins equivalent to the multiple of $2\pi$ that must be
subtracted from their difference to give a value in the range
$\pm\pi$. The sum over the winding numbers around a closed path is
then the vorticity of the enclosed region. In a system with $N=3600$
we observe a tightly bound vortex/anti-vortex pair at $T/J=1.5$
(figure~\ref{fig:2dXY_harmonic_vortices}). It appears as though the
vortices are created by the superposition of high energy spin waves, in
contradiction to the assumption that spin waves and vortices are
independent~\cite{kosterlitz_1973_1181,berezinskii_1971_493}
(although, as
previous studies have generally focused on temperatures far below
$T_{KT}$ this should have no bearing on their results). The
vortex density increases with $T$, eventually
becoming so large that it is difficult to distinguish individual
pairs. At no point do any vortices have a vorticity greater than 1
and isolated vortices are never observed.

The lack of vortex pair unbinding can be explained by considering the
free energy of an isolated vortex. Adapting the standard calculation
(see, for example,~\cite{chaikin_book}) to allow for the non-periodic
spins, we arrive at an expression for the energy of a vortex, 
\begin{equation}
E_{\mathrm{vor}} = 2J\pi^2(L-a)-J\pi l\ln\left(\frac{L}{a}\right),
\end{equation}
where $l$ is a constant of the order of the lattice spacing $a$. That
$E_{\mathrm{vor}}$ diverges as $\sqrt{N}$ means that it dominates the
entropy (which 
diverges logarithmically with $N$) at all temperatures. Thus isolated
vortices are always energetically unstable in the Harmonic
model. Bound pairs of vortices whose cores are separated by a distance
$R$ will have
\begin{equation}
E_{\mathrm{vor-pair}} \sim \mathcal{O}(R-\ln(R))
\end{equation}
and are therefore feasible provided $R$ remains small.

Despite the observation of topological defects, the evidence from
statistics and the behaviour of the 
susceptibility, the lack of vortex-pair unbinding must lead to the
conclusion that no 
Kosterlitz-Thouless-Berezinski\u{i} transition occurs in the Harmonic
model. 

\section{Conclusion}

In this work we have considered the PDF of order parameter
fluctuations in the finite 2d$XY$ model
in the spin wave approximation. We have 
not generally considered the effect of spin vortices on the PDF. Bound
vortex pairs are irrelevant 
perturbations in the 
low temperature phase and so might be expected to leave the PDF
unaltered, but this has not been proved explicitly.  
For a study of the effect of vortex unbinding on the PDF,
see~\cite{holdsworth_2002_643}. 

The original result of \cite{bramwell_2001_041106} suggested a PDF
describing magnetization fluctuations in critical finite 2d$XY$
magnets which
was truly universal and had been seen to describe the statistics of a
range of critical systems. We have demonstrated analytically that
there is in fact a temperature dependence and so
it must be concluded that strict universality does not hold. However,
numerical work shows that the effect of temperature is very 
weak. From a visual point of view the distribution looks very similar 
across the critical, vortex free, region.
The temperature dependence established here is weaker than, but of the
same magnitude as, that claimed by Palma {\it et al.}, for the full XY
model \cite{palma_2002_026108}. More work is needed to fully
understand this difference.

Despite the weakness of the temperature dependence, we feel that the
confirmation of its existence is an important result, particularly as
much literature has grown up around the unusual behaviour of
fluctuations in the 2d$XY$ model. Significantly it has been shown that
the equivalence between the 
distributions obtained using the cosine and linearized order
parameters in~\cite{bramwell_2001_041106} is the result of the
imposition of low temperature in both 
cases -- by neglect of MLGs and anharmonic terms respectively. 

We note that the assertion that the Harmonic model is `vortex free' is
true only in the sense that vortices do not appear explicitly in the
Hamiltonian. However, by introducing a definition of vortices
consistent with infinitely variable spins, we have identified tightly
bound vortex pairs at high temperatures which we conclude are the
result of the superposition of high energy spin waves. The crossover
to a Gaussian statistical regime at $T\approx4\pi$ for the Harmonic
model, coupled with the fact that the susceptibility appears not to
diverge above this point, are suggestive of a
Kosterlitz-Thouless-Berezinski\u{i} 
transition. However, the unbinding of vortex pairs this requires is
not observed and we argue that isolated vortices in this model are not
energetically viable.

\ack
It is a pleasure to thank P.\ C.\ W.\ Holdsworth, Z.\ R\'acz and M.\
Clusel for useful discussions. S.\ T.\ Banks thanks the UCL Graduate
School for funding through a Graduate School Research Scholarship.

\end{document}

%% file: figs/graphs_a.tex
\begin{picture}(140,30)(0,0)
\put(10,0){\circle*{10}}
\put(10,30){\circle*{10}}
\put(40,0){\circle*{10}}
\put(55,25.981){\circle*{10}}
\put(70,0){\circle*{10}}
\put(100,0){\circle*{10}}
\put(100,30){\circle*{10}}
\put(130,0){\circle*{10}}
\put(130,30){\circle*{10}}
\thicklines
\qbezier(10,0)(0,15)(10,30)
\qbezier(10,0)(20,15)(10,30)
\put(40,0){\line(3,5){15}}
\put(40,0){\line(1,0){30}}
\put(70,0){\line(-3,5){15}}
\put(100,0){\line(0,1){30}}
\put(100,0){\line(1,0){30}}
\put(100,30){\line(1,0){30}}
\put(130,0){\line(0,1){30}}
\put(25,0){\Huge{,}}
\put(85,0){\Huge{,}}
\end{picture}

%% file: figs/graphs_b.tex
\begin{picture}(100,40)(0,0)
\thicklines
\dotty{10}{0}{10}
\dotty{10}{30}{10}
\dotty{40}{15}{10}
\qbezier(10,0)(0,15)(10,30)
\qbezier(10,0)(20,15)(10,30)
\put(10,0){\line(2,1){30}}
\put(10,30){\line(2,-1){30}}
\dotty{70}{0}{10}
\dotty{70}{30}{10}
\dotty{100}{0}{10}
\dotty{100}{30}{10}
\put(70,0){\line(0,1){30}}
\put(70,0){\line(1,0){30}}
\put(100,0){\line(0,1){30}}
\put(70,30){\line(1,0){30}}
\put(107.070,37.070){\circle{20}}
\put(50,0){\Huge{,}}
\end{picture}

%% file: figs/graphs_c.tex
\begin{picture}(110,50)(0,0)
\thicklines
\dotty{15}{10}{10}
\dotty{15}{40}{10}
\dotty{45}{10}{10}
\dotty{45}{40}{10}
\qbezier(15,10)(5,25)(15,40)
\qbezier(15,10)(25,25)(15,40)
\qbezier(45,10)(35,25)(45,40)
\qbezier(45,10)(55,25)(45,40)
\dotty{90}{10}{10}
\dotty{90}{40}{10}
\qbezier(90,10)(80,25)(90,40)
\qbezier(90,10)(100,25)(90,40)
\put(62.5,22.5){\line(1,0){10}}
\put(62.5,25){\line(1,0){10}}
\put(5,25){\oval(5,40)[l]}
\put(55,25){\oval(5,40)[r]}
\put(80,25){\oval(5,40)[l]}
\put(100,25){\oval(5,40)[r]}
\put(107.5,40){\Large{2}}
\end{picture}

%% file: figs/mlgs.tex
\begin{picture}(60,35)(0,0)
\dotty{0}{20}{10}
\dotty{60}{20}{10}
\thicklines
\qbezier(0,20)(30,55)(60,20)
\qbezier(0,20)(30,40)(60,20)
\qbezier(0,20)(30,0)(60,20)
\qbezier(0,20)(30,-15)(60,20)
\end{picture}

%% file: figs/data.tex
\begin{tabular}{ccc}
\scriptsize{
\begin{tabular}{|c|c|c|c|}
\hline
\hline
   {$T$} & {$K_{\mathrm{eff}}$} & {$T_{\mathrm{eff}}$} &    {$y$}	 \\
\hline
\hline
       0.1 &         10 &        0.1 &         0.000\\
\hline
       0.2 &          5 &        0.2 &         0.000\\
\hline
       0.3 &      3.333 &        0.3 &         0.000\\
\hline
       0.4 &        2.5 &        0.4 &         0.000\\
\hline
       0.5 &          2 &        0.5 &         0.000\\
\hline
       0.6 &      1.667 &        0.6 &         0.000\\
\hline
       0.7 &      1.429 &        0.7 &         0.000\\
\hline
       0.8 &       1.25 &        0.8 &         0.000\\
\hline
       0.9 &      1.111 &        0.9 &         0.000\\
\hline
         1 &      0.999 &      1.001 &      0.001 \\
\hline
\end{tabular}}  &
\scriptsize{
\begin{tabular}{|c|c|c|c|}
\hline
\hline
   {$T$} & {$K_{\mathrm{eff}}$} & {$T_{\mathrm{eff}}$} &    {$y$} \\
\hline
\hline
       1.1 &      0.905 &      1.105 &      0.002 \\
\hline
       1.2 &      0.825 &      1.213 &      0.004 \\
\hline
       1.3 &       0.75 &      1.333 &       0.01 \\
\hline
       1.4 &      0.677 &      1.478 &       0.02 \\
\hline
       1.5 &      0.596 &      1.679 &      0.038 \\
\hline
       1.6 &        0.5 &          2 &      0.071 \\
\hline
       1.7 &      0.386 &      2.592 &      0.129 \\
\hline
       1.8 &      0.263 &      3.804 &      0.232 \\
\hline
       1.9 &      0.157 &      6.388 &      0.401 \\
\hline
         2 &      0.085 &      11.74 &      0.655 \\
\hline
\end{tabular}} 
\end{tabular}

%% file: 2dxy.bbl
\section*{References}
\begin{thebibliography}{99}
\bibitem{landau_book} Landau L D and Lifshitz E M 1980 {\it
Statistical Physics} vol 1 (Pergamon)
\bibitem{wilson_1974_75} Wilson K G and Kogut J 1974 \emph{Phys.\ Rep.}
{\bf 12} 75
\bibitem{cassandro_1978_913} Cassandro M and Jona-Lasinio G 1978
\emph{Adv.\ Phys.} {\bf 27} 913
\bibitem{garrod_book} Garrod C 1995 {\it Statistical Mechanics and
Thermodynamics} (Oxford University Press)
\bibitem{cardy_book} Cardy J 1996 {\it Scaling and Renormalization in
Statistical Physics} (Cambridge University Press)
\bibitem{bruce_1981_3667} Bruce K 1981 \emph{J.\ Phys.\ C} {\bf 14} 3667
\bibitem{binder_inbook} Binder K P 1992 in {\it Computational Methods in
Field Theory -- Lecture Notes in Physics} vol 490 ed H Gausterer and C
B Lang (Springer-Verlag) 
\bibitem{botet_2000_1825} Botet R and Ploszajczak M 2000
\emph{Phys.\ Rev.\ E} {\bf 62} 1825
\bibitem{bramwell_1998_552} Bramwell S T, Holdsworth P C W and Pinton
J-F 1998 \emph{Nature} {\bf 396} 552
\bibitem{bramwell_2000_3744} Bramwell S T, Christensen K, Fortin J-Y,
Holdsworth P C W, Jensen H J, Lise S, L\'opez J, Nicodemi M, Pinton
J-F and Sellitto M 2000 \PRL {\bf 84} 3744
\bibitem{kosterlitz_1974_1046} Kosterlitz J M 1974 \JPC {\bf 7} 1046
\bibitem{archambault_1997_8363} Archambault P, Bramwell S T and
Holdsworth P C W 1997 \JPA {\bf 30} 8363
\bibitem{bramwell_2001_041106} Bramwell S T, Fortin J-F, Holdsworth P
C W, Peysson S, Pinton J-F, Portelli B and Sellitto M 2001
\emph{Phys. Rev. E} {\bf 63} 2001
\bibitem{holdsworth_2002_643} Holdsworth P C W and Sellitto M 2002
\emph{Physica A} {\bf 315} 643
\bibitem{portelli_2003_104501} Portelli B, Holdsworth P C W and Pinton
J-F 2003 \PRL {\bf 90} 104501
\bibitem{peyrard_2004_265} Peyrard M 2004 \emph{Physica D} {\bf 193}
265
\bibitem{peyrard_2002_834} Peyrard M and Daumont I 2002
\emph{Europhys.\ Lett.} {\bf 59} 834
\bibitem{noullez_2002_231} Noullez A and Pinton J-F 2002
\emph{Eur.\ Phys.\ J.\ B} {\bf 28} 231
\bibitem{tothkatona_2004_016302} Toth-Katona T and Gleeson J T 2004
\emph{Phys.\ Rev.\ E} {\bf 69} 016302
\bibitem{tothkatona_2003_264501} Toth-Katona T and Gleeson J T 2003
\PRL {\bf 91} 264501
\bibitem{goldburg_2001_245502} Goldburg W I, Goldschmidt Y Y and
Kellay H 2001 \PRL {\bf 87} 245502
\bibitem{bramwell_2002_310} Bramwell S T, Fennel T, Holdsworth P C W
and Portelli B 2002 \emph{Europhys.\ Lett.} {\bf 57} 310
\bibitem{pennetta_2004_s164} Pennetta C, Alfinito E, Reggiani L and
Ruffo S 2004 \SST {\bf 19} S164
\bibitem{pennetta_2004_380} Pennetta C, Alfinito E, Reggiani L and
Ruffo S 2004 \emph{Physica A} {\bf 340} 380
\bibitem{sinharay_2001_186} Sinha-Ray P, de Agua L B and Jensen H J
2001 \emph{Physica D} {\bf 157} 186
\bibitem{dahlstedt_2001_11193} Dahlstedt K and Jensen H J 2001 \JPA
{\bf 34} 11193
\bibitem{chapman_2002_409} Chapman S C, Rowlands G and Watkins N W
2003 \emph{Nonlin.\ Proc.\ Geoph.} {\bf 9} 409
\bibitem{chamon_2004_10120} Chamon C, Charbonneau P, Cugliandolo L
F, Reichman D R and Sellitto M 2004 \JCP {\bf 121} 10120
\bibitem{clusel_2004_046112} Clusel M, Fortin J-Y and Holdsworth P C W
2004 \emph{Phys.\ Rev.\ E} 046112
\bibitem{zheng_2003_026114} Zheng B 2003 \emph{Phys.\ Rev.\ E} {\bf 67}
026114
\bibitem{antal_2001_240601} Antal T, Droz M, Gy\"{o}gyi G and R\'acz Z
2001 \PRL {\bf 87} 240601
\bibitem{racz_1994_3530} R\'acz Z and Plischke M 1994 \emph{Phys.\ Rev.\
E} {\bf 50} 3530
\bibitem{palma_2002_026108} Palma G, Meyer T and Labb\'e R 2002
\emph{Phys.\ Rev.\ E} {\bf 66} 026108
\bibitem{binder_1981_119} Binder K 1981 \emph{Z.\ Phys.\ B} {\bf 43} 119
\bibitem{berezinskii_1971_493} Berezinski\u{i} V L 1971
\emph{Sov.\ Phys.\ JETP} {\bf 32} 493
\bibitem{villain_1975_581} Villain J 1975 \emph{J.\ Physique} {\bf 36}
581
\bibitem{mermin_1966_1133} Mermin N D and Wagner H 1966 \PRL {\bf 17}
1133
\bibitem{bramwell_1993_L53} Bramwell S T and Holdsworth P C W 1993
\JPCM {\bf 5} L53
\bibitem{bramwell_1994_8811} Bramwell S T and Holdsworth P C W 1994
\emph{Phys.\ Rev.\ B} {\bf 49} 8811
\bibitem{jose_1977_1217} Jos\'e J V, Kadanoff L P, Kirkpatrick S and
Nelson D R 1977 \emph{Phys.\ Rev.\ B} {\bf 16} 1217
\bibitem{kendall_book} Kendall M, Stuart A and Ord J K 1987 {\it
Kendall's  Advanced Theory of Statistics} vol 1 (London: Griffin)
\bibitem{archambault_1998_7234} Archambault P, Bramwell S T, Fortin
J-Y, Holdsworth P C W, Peysson S and Pinton J-F 1998 \emph{J.\ App.\
Phys.} {\bf 83} 7234 
\bibitem{metropolis_1953_1087} Metropolis N, Rosenbluth A, Rosenbluth
M, Teller A and Teller E 1953 \JCP {\bf 21} 1087
\bibitem{kosterlitz_1973_1181} Kosterlitz J M and Thouless D J 1973
\JPC {\bf 6} 1181
\bibitem{chaikin_book} Chaikin P M and Lubensky T C 1995 {\it
Principles of Condensed Matter Physics} (Cambridge University Press)
\end{thebibliography}
